\documentclass[12pt]{article}
\usepackage[utf8]{inputenc}
\usepackage{amsmath,amssymb,aas_macros}
\usepackage{graphicx}
\usepackage{color}
\usepackage{cite}
\usepackage{mathrsfs}
\usepackage{bm}
\usepackage{enumerate}

\usepackage[x11names]{xcolor}

\usepackage[colorlinks,citecolor=blue]{hyperref}
\renewcommand{\thefootnote}{\#\arabic{footnote}}

\makeatletter
\@addtoreset{equation}{section}
\makeatother

\begin{document}

\begin{titlepage}

\begin{center}

\vskip .75in

{\Large \bf Probing isocurvature perturbations 
\vspace{2mm} \\ with 21-cm global signal in the light of HERA result}

\vskip .75in

{\large
Teppei Minoda$\,^{a,b}$, Shintaro Yoshiura$\,^{c,a}$, Tomo Takahashi$\,^d$
}

\vskip 0.25in

{\em
$^{a}$The University of Melbourne, School of Physics, Parkville, VIC 3010, Australia
\vspace{2mm} \\
$^{b}$Department of Physics and Astrophysics, Nagoya University, Chikusa-ku, Nagoya, 464-8602, Japan
\vspace{2mm} \\
$^{c}$Mizusawa VLBI Observatory, National Astronomical Observatory Japan, 2-21-1 Osawa, Mitaka, Tokyo 181-8588, Japan
\vspace{2mm} \\
$^{d}$Department of Physics, Saga University, Saga 840-8502, Japan 
}

\end{center}
\vskip .5in

\begin{abstract}
We argue that the 21-cm global signal can be a powerful probe of isocurvature perturbations, particularly for the ones with blue-tilted spectra. Although the 21-cm global signal is much affected by astrophysical processes, which give some uncertainties when cosmological models are investigated, recent results from HERA have constrained several astrophysical parameters, whose information can reduce the ambiguities originating from astrophysics. We show that the size and spectral tilt of isocurvature perturbations can be well inferred from the 21-cm global signal once the information on astrophysics from the HERA results is taken into account. 

\end{abstract}

\end{titlepage}

\renewcommand{\thepage}{\arabic{page}}
\setcounter{page}{1}
\renewcommand{\thefootnote}{\#\arabic{footnote}}
\setcounter{footnote}{0}

\section{Introduction \label{sec:intro}}

Primordial fluctuations are now well measured by observations of the cosmic microwave background (CMB) such as Planck \cite{Planck:2018jri}, which indicates that the primordial power spectrum for the curvature perturbation is almost scale-invariant, but slightly red-tilted, and has an almost Gaussian distribution. Furthermore, the data from Planck also implies that primordial fluctuations are almost adiabatic, which severely constrains some scenarios of the early Universe generating isocurvature fluctuations. Since isocurvature fluctuations are related to the generation mechanism of dark matter (DM) and baryogenesis, they would give important information to probe the phenomena in the early Universe.

However, observations of CMB can only probe fluctuations on large scales. Indeed some scenarios of the early Universe predict isocurvature fluctuations with a blue-tilted spectrum which includes axion \cite{Kasuya:2009up,Chung:2015pga,Chung:2017uzc,Chung:2021lfg}, and primordial black holes (PBHs) \cite{Afshordi:2003zb,Kashlinsky:2016sdv,Gong:2017sie,Gong:2018sos,2019PhRvD.100d3540M,2021PhRvD.104f3522T}.
In this kind of models, isocurvature fluctuations can give a sizable contribution only on small scales whereas they scarcely affect large-scale fluctuations which are severely constrained by CMB, and hence it is imperative to test those scenarios by observations of small scales fluctuations. Actually, there have been several works along this line such as using fluctuations of hydrogen 21-cm line from the intergalactic medium \cite{Kawasaki:2011ze} and minihalos \cite{Takeuchi:2013hza,Sekiguchi:2013lma,Gong:2017sie,Kadota:2020ybe}. Currently, a lot of projects are in progress to detect the 21-cm line signal in the Epoch of Reionization (EoR). Although 21-cm fluctuations from the EoR signal have not been detected yet due to the strong foreground emissions, upper limits on the 21-cm power spectrum have been obtained by the Precision Array for Probing the Epoch of Reionization (PAPER)~\cite{2018ApJ...868...26C,2019ApJ...883..133K}, the Owens Valley Radio Observatory Long Wavelength Array (OVRO-LWA)~\cite{2019AJ....158...84E}, the Low-Frequency Array (LOFAR)~\cite{2020MNRAS.493.1662M}, the Murchison Widefield Array (MWA)~\cite{2016MNRAS.460.4320E,2020MNRAS.493.4711T,2021MNRAS.505.4775Y}, and the Hydrogen Epoch of Reionization Array (HERA)~\cite{2021arXiv210802263T,2021arXiv210807282T}. Furthermore future observations such as the Square Kilometre Array Observatory (SKAO)
\footnote{
{\tt https://www.skatelescope.org}
} are expected to probe 21-cm fluctuations with much better sensitivity.

In this paper, we propose yet another probe of such isocurvature fluctuations: the 21-cm global signal. When small-scale fluctuations are enhanced by the existence of isocurvature perturbations, the structure formation starts from early on, which switches on the sources of {Lyman-$\alpha$} radiation earlier, and then affects the evolution of the spin temperature. Thus not only fluctuations of the 21-cm signal, but also its sky-averaged, or global signal can also probe small-scale fluctuations. Indeed the first detection of absorption signal has been reported by the Experiment to Detect the Global EoR Signature (EDGES) collaboration \cite{Bowman:2018yin} in 2018, which has stimulated a lot of works, including the ones utilizing the detected signal to constrain primordial power spectrum \cite{Yoshiura:2018zts,Yoshiura:2019zxq}, primordial magnetic fields \cite{Minoda:2018gxj,2020MNRAS.498..918B,2021MNRAS.507.1254K}, warm dark matter \cite{Safarzadeh:2018hhg}, decay and annihilation of dark matter \cite{DAmico:2018sxd,Clark:2018ghm,Cheung:2018vww,Liu:2018uzy}, PBHs \cite{Clark:2018ghm,Hektor:2018qqw}, and so on.
On the other hand, there have been some arguments regarding the result  \cite{Hills:2018vyr,2018ApJ...858L..10D,Bradley:2018eev,Singh:2019gsv,Spinelli:2019oqm,2022NatAs.tmp...47S}, and in particular, it has been claimed that recent results from the Shaped Antenna measurement of the background RAdio Spectrum~3 (SARAS 3) experiment excluded the EDGES signal at almost 2$\sigma$ level \cite{2022NatAs.tmp...47S}. Although the EDGES signal would be further carefully investigated and verified, the series of works have shown the power of the 21-cm global signal as a tool to investigate various aspects of cosmology.

In this paper, we investigate how one can apply the 21-cm global signal to probe the amplitude and the scale dependence of isocurvature fluctuations, focusing on cold dark matter (CDM) mode, particularly in the light of the recent HERA result. Indeed the 21-cm line signal is affected not only by cosmological models but also by astrophysical processes, which give degeneracy with cosmological effects such as the one from isocurvature fluctuations. Actually, the HERA collaboration has put constraints on the astrophysical parameters which control the {Lyman-$\alpha$} flux, X-ray heating, and ionizing photons \cite{2021arXiv210807282T}, by combining Thomson scattering optical depth to the CMB, QSOs dark fraction analysis and UV luminosity function of high-$z$ galaxies and upper limits on the 21-cm power spectrum provided by HERA Phase I data \cite{2021arXiv210802263T}.
To evaluate the constraints on the isocurvature fluctuations including astrophysical uncertainties, we adopt the astrophysical models in the numerical simulation \texttt{21cmFAST} \cite{2011MNRAS.411..955M} to calculate the 21-cm global signals, and consider different sets of astrophysical parameters in {the} range of the HERA Phase I constraints \cite{2021arXiv210807282T}.
{We emphasize that the importance and novelty of this work is, to discuss the constraints on the blue-tilted isocurvature perturbations with on the actual observational data: the global absorption signal with EDGES and the upper limit on the 21-cm power spectrum with HERA.}

The structure of this paper is as follows. In the next section, we summarize how we calculate matter power spectrum in models with CDM isocurvature fluctuations to study constraints on their properties. {Then,} in Section~\ref{sec:global}, we discuss the 21cm global signal with the existence of blue-tilted isocurvature fluctuations, including astrophysical uncertainties. Constraints on CDM isocurvature fluctuations are also presented. The final section is devoted to the conclusion of this paper. In this work, we adopt a flat $\Lambda$CDM cosmology, with the cosmological parameters of the matter energy density $\Omega_\mathrm{m}=0.316$, the baryon energy density $\Omega_\mathrm{b}=0.049$, the spectral index of primordial (adiabatic) power spectrum $n_\mathrm{s}^\mathrm{adi}=0.959$, the reduced Hubble constant $h=0.673$, and the amplitude of matter fluctuations in $8~h^{-1} \mathrm{Mpc}$ sphere $\sigma_8=0.811$ which are given by Planck observation \cite{2020A&A...641A...6P}.

\section{CDM isocurvature perturbations and matter power spectrum
\label{sec:power}}
The initial conditions for the adiabatic and CDM isocurvature modes are characterized by the curvature perturbations $\zeta$ and the entropy perturbations $S_\mathrm{CDM}$, respectively. We assume the power-law form for the power spectra as
\begin{align}
\label{eq:power_adi}
\mathcal{P}_\zeta (k) = A_\mathrm{s}^\mathrm{adi} \left(\cfrac{k}{k_*}\right)^{n_\mathrm{s}^\mathrm{adi}-1}~, \\
\label{eq:power_iso}
\mathcal{P}_{S_\mathrm{CDM}} (k) = A^\mathrm{iso} \left(\cfrac{k}{k_*}\right)^{n^\mathrm{iso}-1}~,
\end{align}
where $A_{\rm s}^{\rm adi}$ and $A^{\rm iso}$ are the amplitudes for the adiabatic and isocurvature modes at the pivot scale $k_\ast$ and $n_s^{\rm adi}$ and $n^{\rm iso}$ are their spectral indices. We take the pivot scale as $k_* \equiv 0.05~\mathrm{Mpc}^{-1}$, which is often used in the CMB data analysis. To parametrize the amplitude of the isocurvature perturbations $A^\mathrm{iso}$, we introduce the fraction parameter with
\begin{align}
\label{eq:iso_fraction}
r_\mathrm{CDM} = \cfrac{A^\mathrm{iso}}{A_s^\mathrm{adi}}~.
\end{align}
Current constraints on the uncorrelated CDM isocurvature perturbations give $r_\mathrm{CDM}<\mathcal{O}(0.01)$ \cite{Planck:2018jri}. To calculate the matter power spectrum, we need the transfer functions for the adiabatic and isocurvature modes. In this work, we take the BBKS expression~\cite{1986ApJ...304...15B} and its modification~\cite{1995ApJS..100..281S} as
\begin{align}
T_{\mathrm{adi}}(k)&=\cfrac{\ln (1+2.34 q)}{2.34 q} \times
\left[1+3.89 q+(16.1 q)^{2}+(5.46 q)^{3}+(6.71 q)^{4}\right]^{-1/4} \\
T_{\mathrm{iso}}(k)&= 
{\left[1+\cfrac{(40 q)^{2}}{1+215 q+(16 q)^{2}(1+0.5 q)^{-1}}+(5.6 q)^{8 / 5}\right]^{-5/4},}
\end{align}
where $q=k/(\Omega_\mathrm{m}h^2 \exp(\Omega_\mathrm{b}-\Omega_\mathrm{b}/\Omega_\mathrm{m})~[\mathrm{Mpc}^{-1}])$.
Then, the total matter power spectrum is given by
\begin{align}
P_\mathrm{m}(k) &= 
\left[\mathcal{P}_\zeta(k) T_\mathrm{adi}^2(k)
+ \mathcal{P}_{S_\mathrm{CDM}}(k) T_\mathrm{iso}^2(k)\right] \nonumber \\
&= A_\mathrm{s}^\mathrm{adi} \left(\frac{k}{k_*}\right)^{n_\mathrm{s}^\mathrm{adi}-1}
\left[T_\mathrm{adi}^2(k)
+r_\mathrm{CDM} \left(\frac{k}{k_*}\right)^{n^\mathrm{iso}-n_\mathrm{s}^\mathrm{adi}}
 T_\mathrm{iso}^2(k)\right]~.
 \label{eq:power_tot}
\end{align}
In the following analysis, we consider uncorrelated isocurvature perturbations as shown in Eq. \eqref{eq:power_tot}, which arises in CDM axion models. Although such models generally predict $n^\mathrm{iso} \sim 1$, blue-tilted spectra can also be predicted in some scenarios \cite{Kasuya:2009up,Chung:2015pga,Chung:2017uzc,Chung:2021lfg}. Therefore we take $n^\mathrm{iso}$ as a free parameter as well as $r_\mathrm{CDM}$ to characterize the statistical property of the isocurvature perturbations.

In Eq. \eqref{eq:power_tot}, we determine $A_\mathrm{s}^\mathrm{adi}$ by imposing $\sigma_8 = 0.811$. Since CDM isocurvature fluctuations give a subdominant contribution on large scales, in most parameter spaces we consider in this paper, the amplitude of the large-scale power spectrum is {the} same as the one with the adiabatic mode only. When both $r_{\rm CDM}$ and $n^{\rm iso}$ are large, the large-scale amplitude of the power spectrum should get suppressed by several percent to have $\sigma_8 =0.811$, due to the contribution from the isocurvature mode, however, the change scarcely affects our arguments.

\begin{figure}
\centering
\includegraphics[width=10cm]{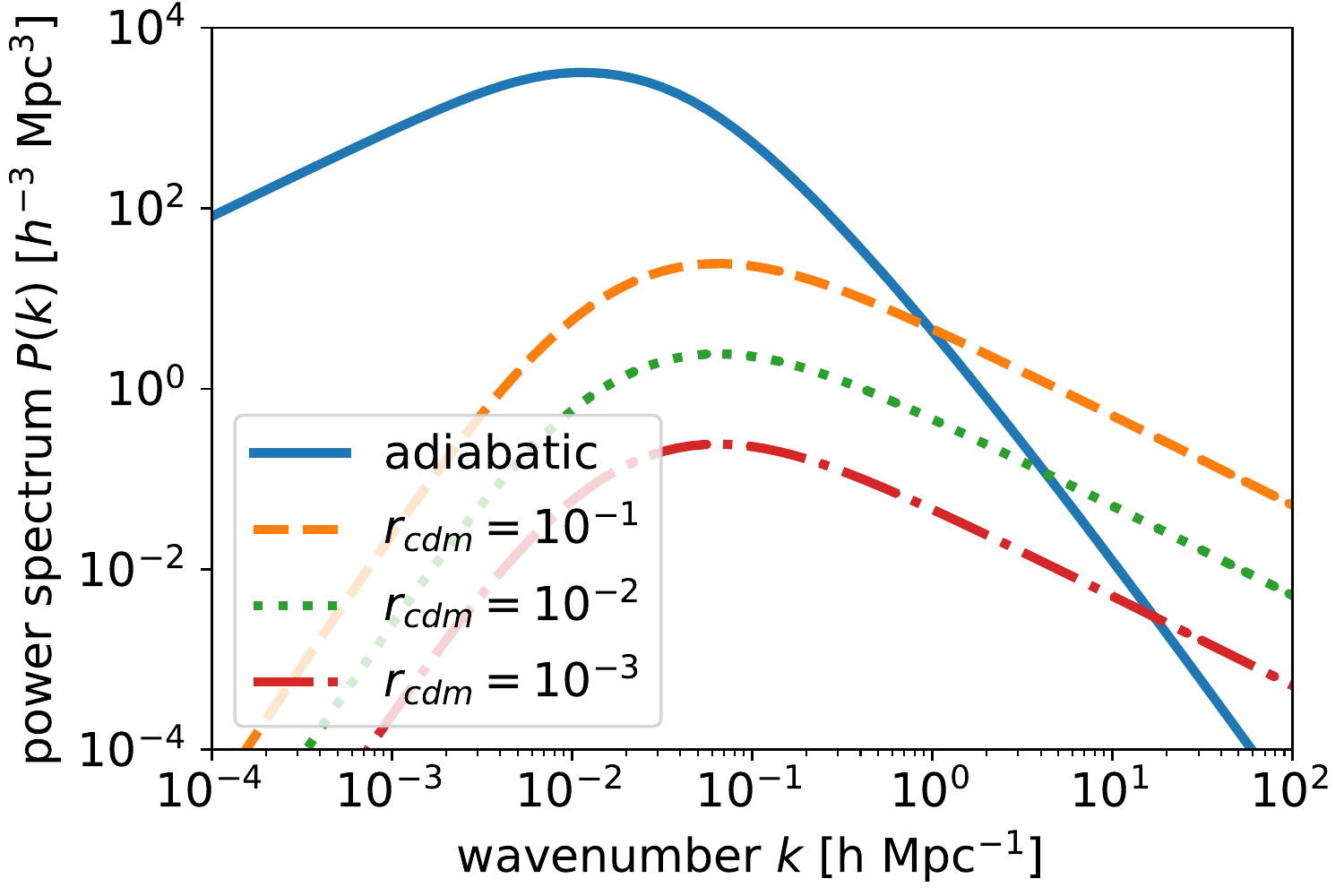}
\caption{
The matter power spectra for adiabatic and CDM isocurvature mode perturbations.
Here, we fix $n^\mathrm{iso}=3.0$.
}
\label{fig:matterpower}
\end{figure}

Figure~\ref{fig:matterpower} shows the matter power spectra for the cases with adiabatic perturbations alone, isocurvature perturbations with $n^\mathrm{iso}=3.0$ and $r_\mathrm{CDM}=10^{-3}, 10^{-2},$ and $10^{-1}$. While the contribution from isocurvature fluctuations is subdominant on large scales  $k<1~h~\mathrm{Mpc}^{-1}$, small scale fluctuations are much affected by isocurvature fluctuations as seen from Figure~\ref{fig:matterpower}. Thus it is expected that small-scale or high-redshift structure formation is a good probe for the isocurvature perturbations. 

\section{21-cm global signal as a probe of isocurvature perturbations
\label{sec:global}}
Now in this section, we discuss how the 21-cm global signal is affected by varying $r_\mathrm{CDM}$ and $n^\mathrm{iso}$ which are defined in Eqs.~\eqref{eq:power_iso} and \eqref{eq:iso_fraction}. The intensity of the 21-cm line observed at a frequency $\nu=1.42~\mathrm{GHz}/(1+z)$ is conventionally represented by the differential brightness temperature $\delta T_\mathrm{b}(\nu)$ and approximately given by \cite{1997ApJ...475..429M}
\begin{align}
\delta T_\mathrm{b}(\nu) \simeq 27 x_\mathrm{HI}(z)
\left(\cfrac{\Omega_\mathrm{b}h^2}{0.023}\right)
\left(\cfrac{0.15}{\Omega_\mathrm{m}h^2}\right)^{1/2}
\left(\cfrac{1+z}{10}\right)^{1/2}
\left(1-\cfrac{T_\mathrm{rad}(z)}{T_\mathrm{S}(z)}\right)~[\mathrm{mK}].
\label{eq:deltaTb}
\end{align}
Here $T_\mathrm{S}$, $T_\mathrm{rad}$, and $x_\mathrm{HI}$ are the neutral hydrogen spin temperature, the background radiation temperature, and the averaged neutral fraction of hydrogen, respectively. From Eq.~\eqref{eq:deltaTb}, the 21-cm global signal can be observed as an emission one when $T_\mathrm{rad}<T_\mathrm{S}$, whereas as an absorption one  when $T_\mathrm{rad} > T_\mathrm{S}$. In this work, since we simply assume that the background radiation temperature is the CMB one, $T_\mathrm{rad}$ is given as $T_\mathrm{CMB}\simeq 2.73\, (1+z)~\mathrm{K}$. Furthermore, the neutral fraction {of hydrogen} should be unity at $z>10$ not to violate the constraint on the Thomson scattering optical depth to the CMB. Therefore the frequency dependence of the global signal is dominantly determined by the redshift evolution of the neutral hydrogen spin temperature at $z>10$.

There are three main physical effects to determine the spin temperature of neutral hydrogen, i.e., collisions with hydrogen themselves, collisions between the hydrogen and CMB photons, and interactions with the background {Lyman-$\alpha$} photons. In particular, {Lyman-$\alpha$} photons produced by the first-generation stars are considered to couple the spin temperature with the kinetic gas temperature. Because of this {Lyman-$\alpha$} coupling, an absorption signal at the frequency corresponding to some high redshift is expected although the amplitude of the signal and its position in the frequency space depend on models, which indicates that measurements of the 21-cm global signal can serve as a probe of the early structure formation history. On the other hand, however, when one tries to investigate cosmological models such as the one including isocurvature fluctuations using the 21-cm signal, the effects of the above-mentioned astrophysical processes are regarded as uncertainties. Therefore information or constraints on such astrophysics would be helpful to probe cosmological models in this regard. Indeed recently HERA phase I results gave constraints on astrophysical models \cite{2021arXiv210807282T}. In this paper, we take account of the HERA constraints to study constraints on isocurvature fluctuations.

\subsection{Astrophysical uncertainties of the global signal with HERA constraints}
\label{sec:astromodels}
In this work, we calculate the 21-cm line signal with the galaxy-driven reionization model in the numerical code \texttt{21cmFAST} \cite{2011MNRAS.411..955M}. While the \texttt{21cmFAST} astrophysical model has various free parameters, we choose their fiducial values as the mean ones in the recent HERA constraints \cite{2021arXiv210807282T}. To discuss the robustness of our constraints on the isocurvature fluctuations, we also consider different combinations of astrophysical parameters within the HERA constraints. In this subsection, first, we introduce the astrophysical parameters in this model and discuss their impact on the 21-cm global signal. (For detailed descriptions of the model, see Ref.~\cite{2019MNRAS.484..933P}.)

To estimate the strength of {Lyman-$\alpha$} coupling, the UV luminosity function is computed as
\begin{align}
\phi(M_\mathrm{UV}) = \left(f_\mathrm{duty} \cfrac{dn}{dM_\mathrm{h}}\right)
\left|\cfrac{dM_\mathrm{h}}{dM_\mathrm{UV}}\right|~,
\end{align}
where $f_\mathrm{duty}$ is the duty cycle, $dn/dM_\mathrm{h}$ is the halo mass function, and $dM_\mathrm{h}/dM_\mathrm{UV}$ is the halo mass to UV magnitude relation. The duty cycle is parametrized by the minimum halo mass to host galaxies $M_\mathrm{turn}$ as  $f_\mathrm{duty}=\exp{(-M_\mathrm{h}/M_\mathrm{turn})}$, which is motivated by the consideration that the atomic cooling and stellar feedback are ineffective in galaxies with small halo mass. The UV magnitude is estimated from the star formation rate $\Dot{M_*} (M_{\rm h}, z)=M_*/(t_* H(z)^{-1})$, where $t_*$ is the typical star formation timescale normalized by the Hubble time $H(z)^{-1}$ at redshift $z$, and $M_*$ is the stellar mass. $M_*$ is obtained by the power-law stellar-to-halo mass fraction as
\begin{align}
\cfrac{M_*}{M_\mathrm{h}}=f_{*,10}
\left(\cfrac{M_\mathrm{h}}{10^{10}~M_\odot}\right)^{\alpha_*} 
\left(\cfrac{\Omega_\mathrm{b}}{\Omega_\mathrm{m}}\right)~.
\end{align}
The normalization factor is fixed as $f_{*,10}=0.0575$ as the mean value in the HERA constraint \cite{2021arXiv210807282T}, whereas the power-law index $\alpha_*$ is taken to be a free parameter. We fix $f_{*,10}$ because $f_{*,10}$ is well constrained in the HERA analysis, and it strongly degenerates with $t_\ast$.

Not only the {Lyman-$\alpha$} emission, but the X-ray heating and the ionizing photons are also important for the 21-cm global signal calculation. The X-ray heating is determined by the specific X-ray luminosity per unit star formation rate $L_\mathrm{X}/\mathrm{SFR}$, which is given by
\begin{align}
\cfrac{L_{\mathrm{X<2.0keV}}}{\mathrm{SFR}} = \int^{2.0\mathrm{keV}}_{E_0} dE~\cfrac{L_\mathrm{X}}{\mathrm{SFR}}~.
\end{align}
In our calculation, the lowest energy threshold of X-ray heating is fixed as $E_0=0.5~\mathrm{keV}$, and the normalization of $L_\mathrm{X}/\mathrm{SFR}$ is determined by fixing the total X-ray luminosity $L_{\mathrm{X<2.0keV}}/\mathrm{SFR}$, and hence  $L_{\mathrm{X<2.0keV}}/\mathrm{SFR}$ is regarded as a free parameter. For the escape fraction of the ionizing photons, we use $f_\mathrm{esc}(M_\mathrm{h}) = 0.0776({M_\mathrm{h}}/{10^{10}M_\odot})^{0.02}$. We fix $f_\mathrm{esc}$ as we focus on the 21-cm absorption line at $z>10$, and $z_\mathrm{min}$ does not depend on the escape fraction much.

\begin{table}
\begin{center}
\begin{tabular}{|c|c|c|c|c|} 
 \hline
 & $\alpha_*$ & $M_\mathrm{turn}~[M_\odot]$ & $t_*$ & $\log_{10} (L_{\mathrm{X<2.0keV}}/\mathrm{SFR}/[\mathrm{erg~s}^{-1} M_\odot^{-1}~\mathrm{yr}])$ \\
 \hline
 model 1 & 0.50 & $3.8 \times 10^8$ & 0.60 & 40.64\\ 
 \hline
 model 2 & 0.41 & $1.6 \times 10^8$ & 0.29 & 41.52\\ 
 \hline
 model 3 & 0.62 & $1.5 \times 10^9$ & 0.86 & 39.47\\ 
 \hline
\end{tabular}
\caption{
Astrophysical parameters for each model adopted {in Figures \ref{fig:2dConstraint_m1} and \ref{fig:2dConstraint_m23}.}}
\label{tab:astro_params}
\end{center}
\end{table}

To summarize, we vary four parameters $(\alpha_*,~M_\mathrm{turn},~t_*,~L_{\mathrm{X<2.0keV}}/\mathrm{SFR})$ within the HERA Phase I constraint \cite{2021arXiv210807282T} to calculate the 21-cm global signal. The fiducial parameter set corresponds to the mean values of HERA constraints (hereafter we call this parameter set ``model 1''), and we also use two additional parameter sets (model 2 and 3), which produce the 21-cm global signals with the highest and lowest absorption redshifts within 1$\sigma$ HERA constraints and the actual values are shown in Table~\ref{tab:astro_params}. 

We consider models 2 and 3 to show the uncertainties from astrophysics in probing isocurvature perturbations. It should be noted that the parameter constraints from the recent HERA results are not only from 21-cm observation but also from observations of CMB, high-$z$ galaxies, and QSOs. However, the degeneracy between astrophysical parameters and the isocurvature perturbations could still exist, and the HERA's astrophysical parameter constraints would vary by taking into account the isocurvature fluctuations. To verify the degeneracy, one needs to perform parameter analysis which requires high computational costs. The aim of this work is just to investigate how one can probe isocurvature fluctuations with 21-cm global signal observations, given the information on the astrophysical parameters. Therefore, in this work, we only assume that the astrophysical parameters are constrained {to be the} same as {in} \cite{2021arXiv210807282T}.

\begin{figure}
\centering
\includegraphics[width=10cm]{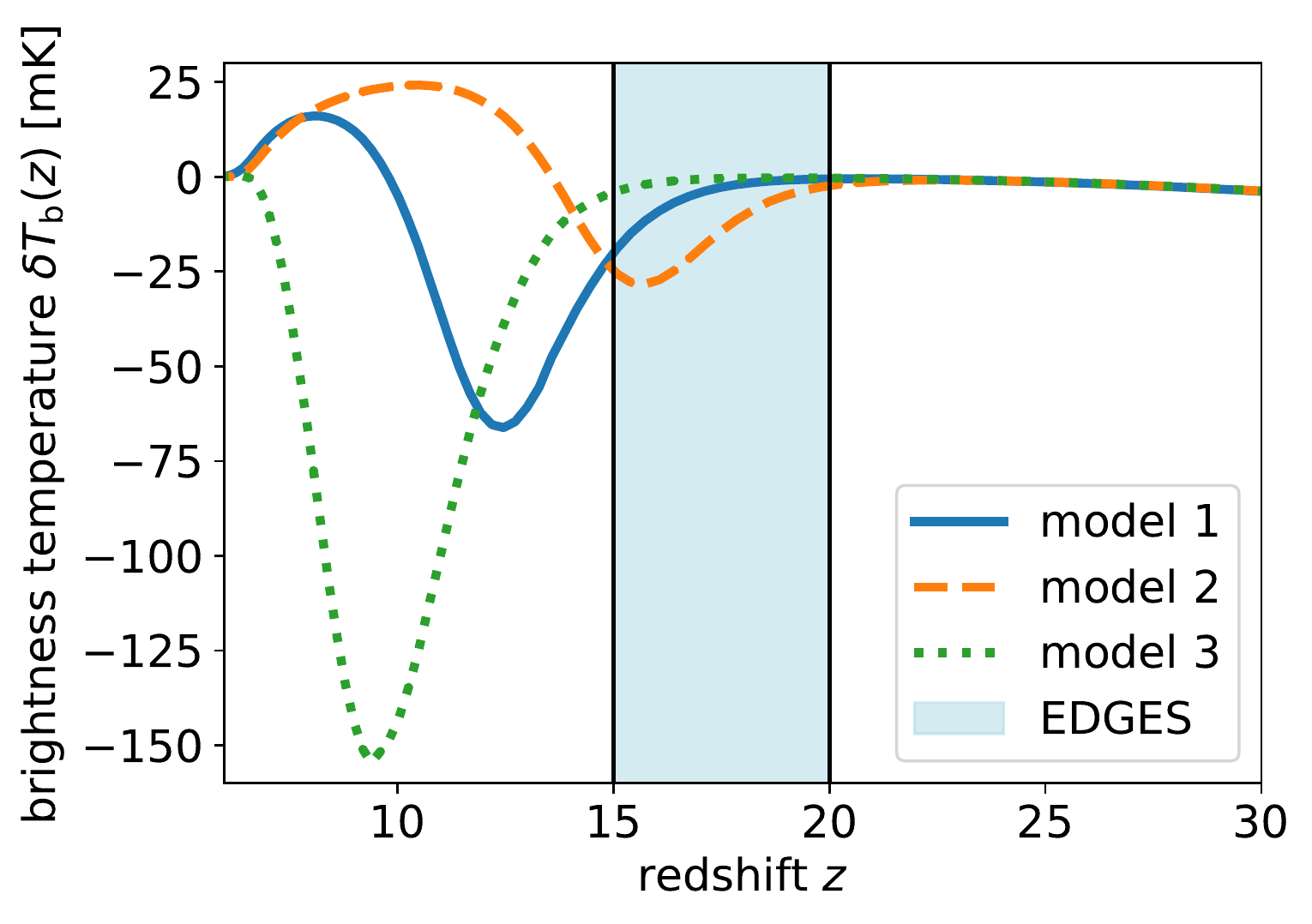}
\caption{
21-cm global signal with different astrophysical models only with the adiabatic fluctuations. The different lines show the calculations with different astrophysical parameter sets, which correspond to Table \ref{tab:astro_params}. The central positions of the absorption profiles are obtained as $z_\mathrm{min}=12.5, 15.7$, and $9.4$ for models 1--3, respectively. To compare with the EDGES obtained signal, we show the redshift range $15<z<20$ with the blue-shaded region.
}
\label{fig:astromodels}
\end{figure}

We plot the 21-cm global signal without isocurvature fluctuations in models 1--3 in Figure \ref{fig:astromodels}. For the following discussions, we define $z_\mathrm{min}$ as the redshift where the brightness temperature takes the lowest value in the absorption trough. We find that $z_\mathrm{min}=12.5, 15.7$, and $9.4$ for models 1--3, respectively. The EDGES observations suggest $15\lesssim z_\mathrm{min} \lesssim 20$, and therefore astrophysical models 1 and 3 are inconsistent with the EDGES data. The parameter $\alpha_*$ is the spectral index of the stellar-to-halo mass relation, and smaller $\alpha_*$ induces the small-scale structure formation. Therefore in model 2 which has the smallest $\alpha_*$, the {Lyman-$\alpha$} coupling is switched on earliest. The parameters $M_\mathrm{turn}$ and $t_*$ also determine the onset of the cosmological star formation, and smaller values for $M_\mathrm{turn}$ and $t_*$ give the higher $z_\mathrm{min}$. In addition, the integrated X-ray luminosity $L_{\mathrm{X<2.0keV}}/\mathrm{SFR}$ determines the redshift when the X-ray heating of the hydrogen gas gets efficient. Since the smaller X-ray luminosity delays the gas heating, the 21-cm global signal results in a deeper absorption and lower $z_\mathrm{min}$.

\subsection{Effects of the isocurvature fluctuations}
In this subsection, we fix the astrophysical parameters with those for model 1 in Table~\ref{tab:astro_params}, and discuss how the 21-cm global signal is affected by the existence of isocurvature fluctuations.

\begin{figure}
\centering
\includegraphics[width=10cm]{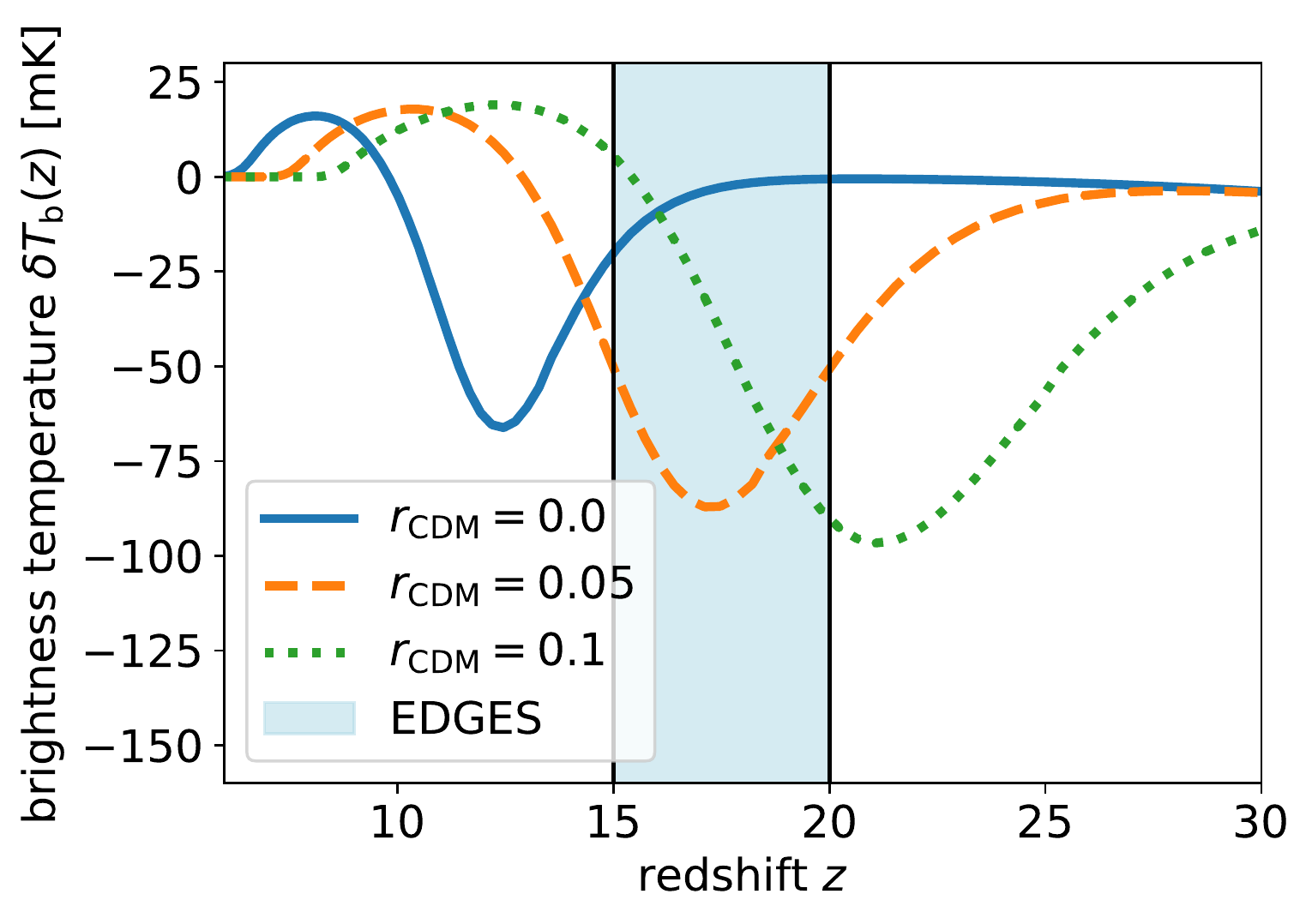}
\caption{
21-cm global signals with different isocurvature fraction  $r_\mathrm{CDM}$. Solid, dashed, and dotted lines correspond to $r_\mathrm{CDM}=0.0,\ 0.05$, and $0.1$, respectively. Here the isocurvature spectral index is fixed to $n^\mathrm{iso}=2.5$. The absorption signal is centered at $z_\mathrm{min}=12.5,\ 17.1,\ 21.1$ for $r_\mathrm{CDM}=0.0,\ 0.05,\ 0.1$, respectively.
}
\label{fig:global_n25}
\end{figure}

First, we fix the isocurvature spectral index as $n^\mathrm{iso}=2.5$, and calculate the 21-cm signal with different isocuvature fractions $r_\mathrm{CDM}=0.0, 0.05,$ and $0.1$ as shown in Figure~\ref{fig:global_n25}. By increasing $r_\mathrm{CDM}$, small-scale power spectrum is enhanced as shown in Section~\ref{sec:power}, and the structure formation at high redshifts are induced. Accordingly, for larger $r_\mathrm{CDM}$, the {Lyman-$\alpha$} coupling between the kinetic gas temperature and the spin temperature becomes effective earlier. As a result, the positions of the absorption signal are obtained as $z_\mathrm{min}=12.5,\ 17.1$, and $21.1$ for $r_\mathrm{CDM}=0.0,\ 0.05$, and $0.1$, respectively. In this model, $r_\mathrm{CDM}=0.0$ and $r_\mathrm{CDM}=0.1$ are rejected if the absorption signal with $15<z_\mathrm{min}<20$ which is suggested by EDGES is confirmed.

\begin{figure}
\centering
\includegraphics[width=10cm]{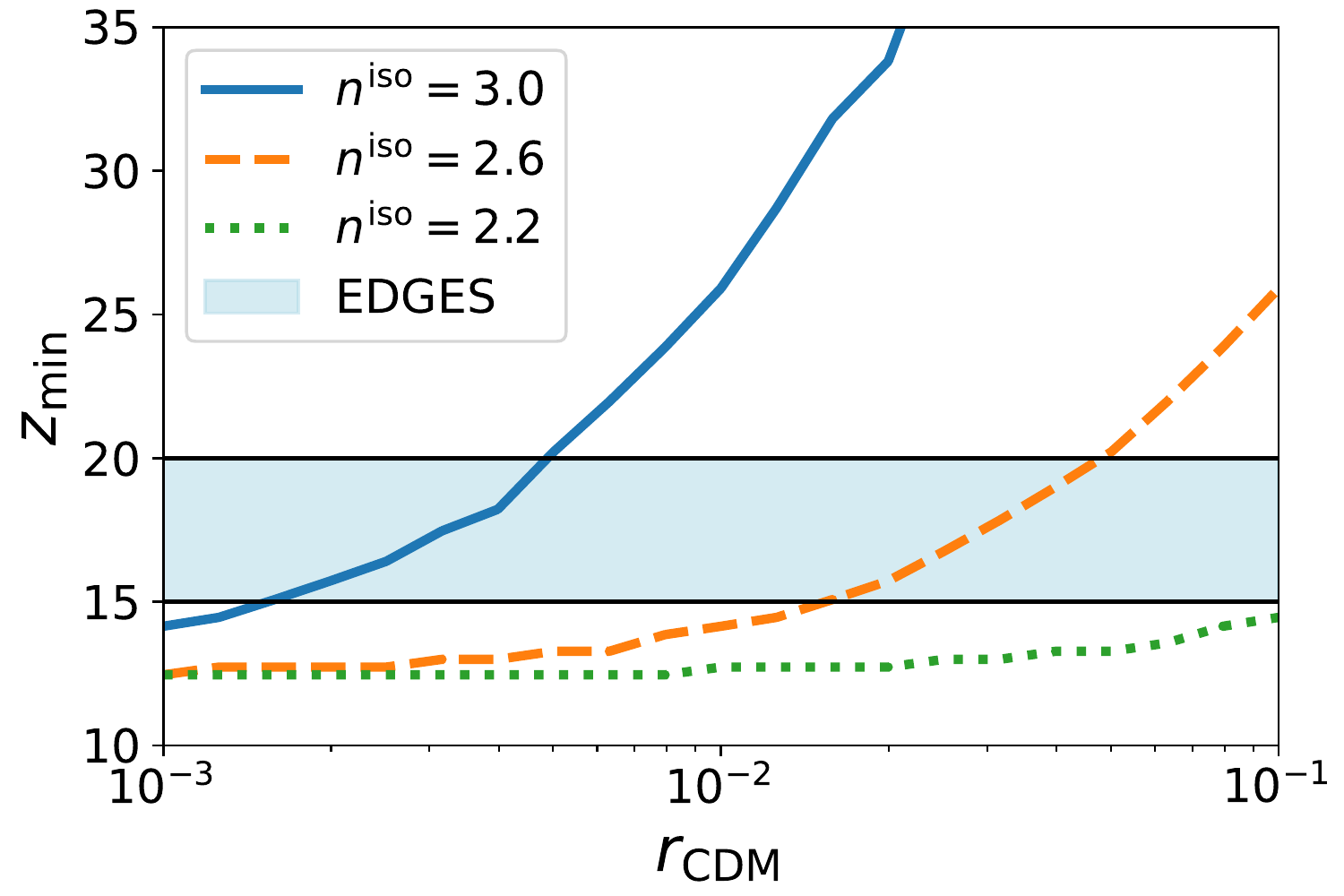}
\caption{
Absorption redshift with varying isocurvature fraction $r_\mathrm{CDM}$ for different isocurvature spectral indices $n^\mathrm{iso}$. Different styles of lines show $n^\mathrm{iso}=2.2$ (dotted), $2.6$ (dashed), and $3.0$ (solid).
}
\label{fig:AbsDepth}
\end{figure}

In Figure~\ref{fig:AbsDepth}, $z_\mathrm{min}$ is plotted as a function of the fraction parameter $r_{\rm CDM}$ for several values of $n^{\rm iso}$. The absorption signal with higher $z_\mathrm{min}$ is obtained by increasing $r_\mathrm{CDM}$ for a fixed $n^\mathrm{iso}$. When the absorption trough in the 21-cm global signal is observationally confirmed, for example, as $15<z_\mathrm{min}<20$ as observed by EDGES, the isocurvature fraction can be constrained as $0.0013 < r_\mathrm{CDM} < 0.005$ for $n^\mathrm{iso}=3.0$ and in this case, non-zero blue-tilted isocurvature fluctuations are favored. As already mentioned, the 21-cm global signal is also affected by astrophysical processes, however, once the information on those parameters is available such as from recent HERA results, our argument shows that the 21-cm global signal can be a very powerful probe of isocurvature fluctuations when the absorption trough is observationally determined. Our results suggest that even the existence of isocurvature mode may be inferred from {the} 21-cm global signal observations given the information on astrophysical parameters.

\subsection{Probing $r_\mathrm{CDM}$ and $n^\mathrm{iso}$ with 21-cm signal}
\label{sec:results}

\begin{figure}
\centering
\includegraphics[width=10cm]{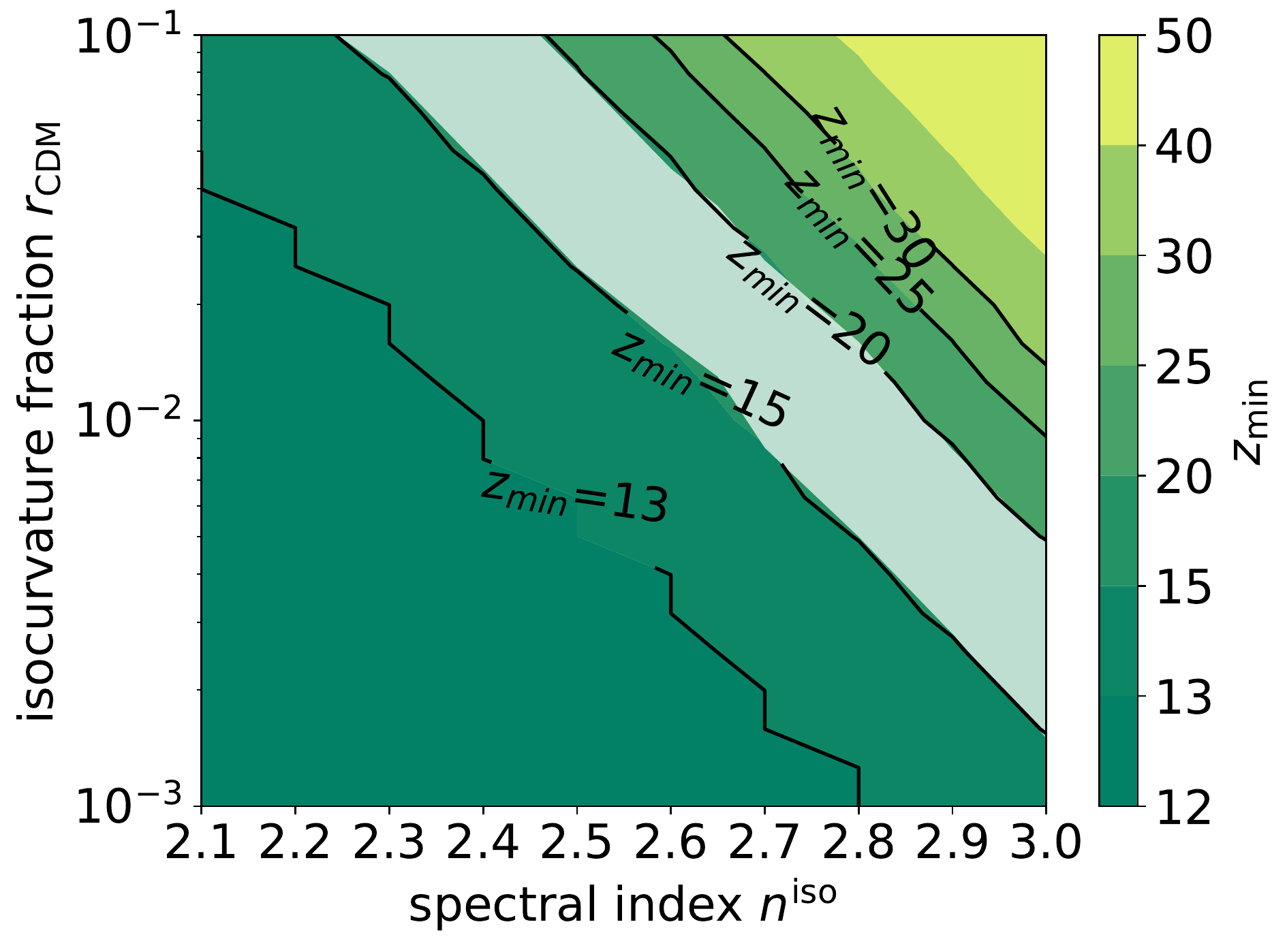}
\caption{
Absorption redshift of the 21-cm global signal in the $n^{\rm iso}$--$r_{\rm CDM}$ plane. The blue-shaded region is consistent with the absorption redshifts suggested by the EDGES experiment $15<z_\mathrm{min}<20$. Here we show the case of astrophysical model 1, in which the mean values of HERA constraints are assumed for astrophysical parameters.}
\label{fig:2dConstraint_m1}
\end{figure}

Figure \ref{fig:2dConstraint_m1} shows the absorption redshifts calculated with various combinations of $(n^\mathrm{iso}, r_\mathrm{CDM})$. When the 21-cm global absorption signal is confirmed in the range of $15<z_\mathrm{min}<20$, the isocurvature fraction can be constrained as $0.0015\lesssim r_\mathrm{CDM}\lesssim 0.005$ for $n^\mathrm{iso}=3.0$, and $0.02\lesssim r_\mathrm{CDM}\lesssim 0.1$ for $n^\mathrm{iso}=2.5$. The left bottom region of the parameter space for $n^\mathrm{iso}$ and $r_\mathrm{CDM}$ is too small to affect the 21-cm global signal, and the resultant $z_\mathrm{min}$ is unchanged.

High values of $n^\mathrm{iso}$ and $r_\mathrm{CDM}$ both enhance small-scale fluctuation and accelerate structure formation at early epochs. Thus, the parameters degenerate each other as shown in Figure \ref{fig:2dConstraint_m1}. For example, we obtain the fitting function for this degeneracy as $\log_{10} r_\mathrm{CDM} \simeq -2.4~ n^\mathrm{iso} +4.4$ if the global signal with $z_\mathrm{min}=15.0$ is confirmed. This degeneracy may be solved by future 21cm line power spectrum observation at large $k$. We leave this issue as future work.

\begin{figure*}
  \begin{minipage}{0.46\linewidth}
    \centering
    \includegraphics[keepaspectratio, scale=0.4]{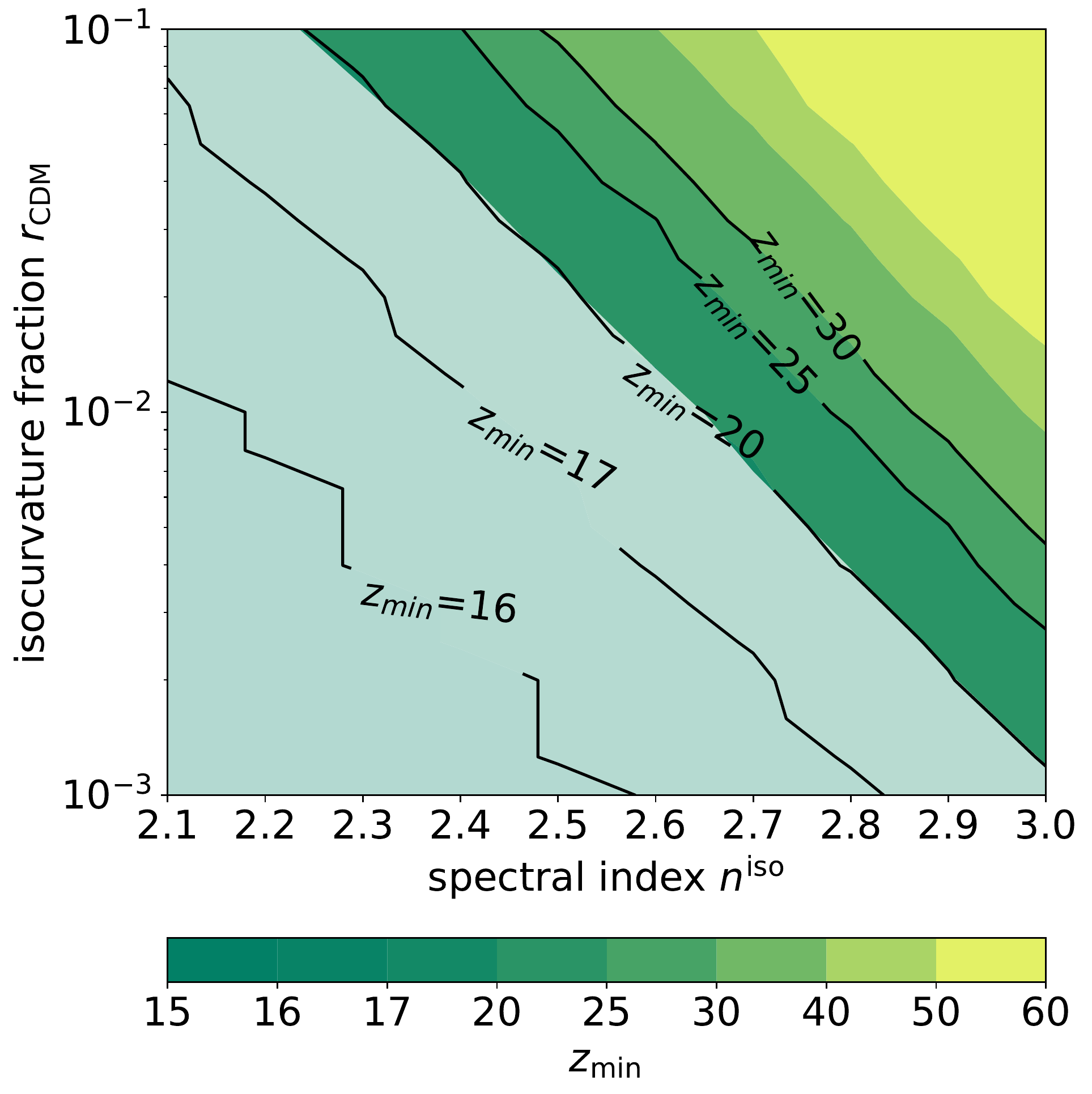}
  \end{minipage}
  \hspace{3mm}
  \begin{minipage}{0.45\linewidth}
    \centering
    \includegraphics[keepaspectratio, scale=0.4]{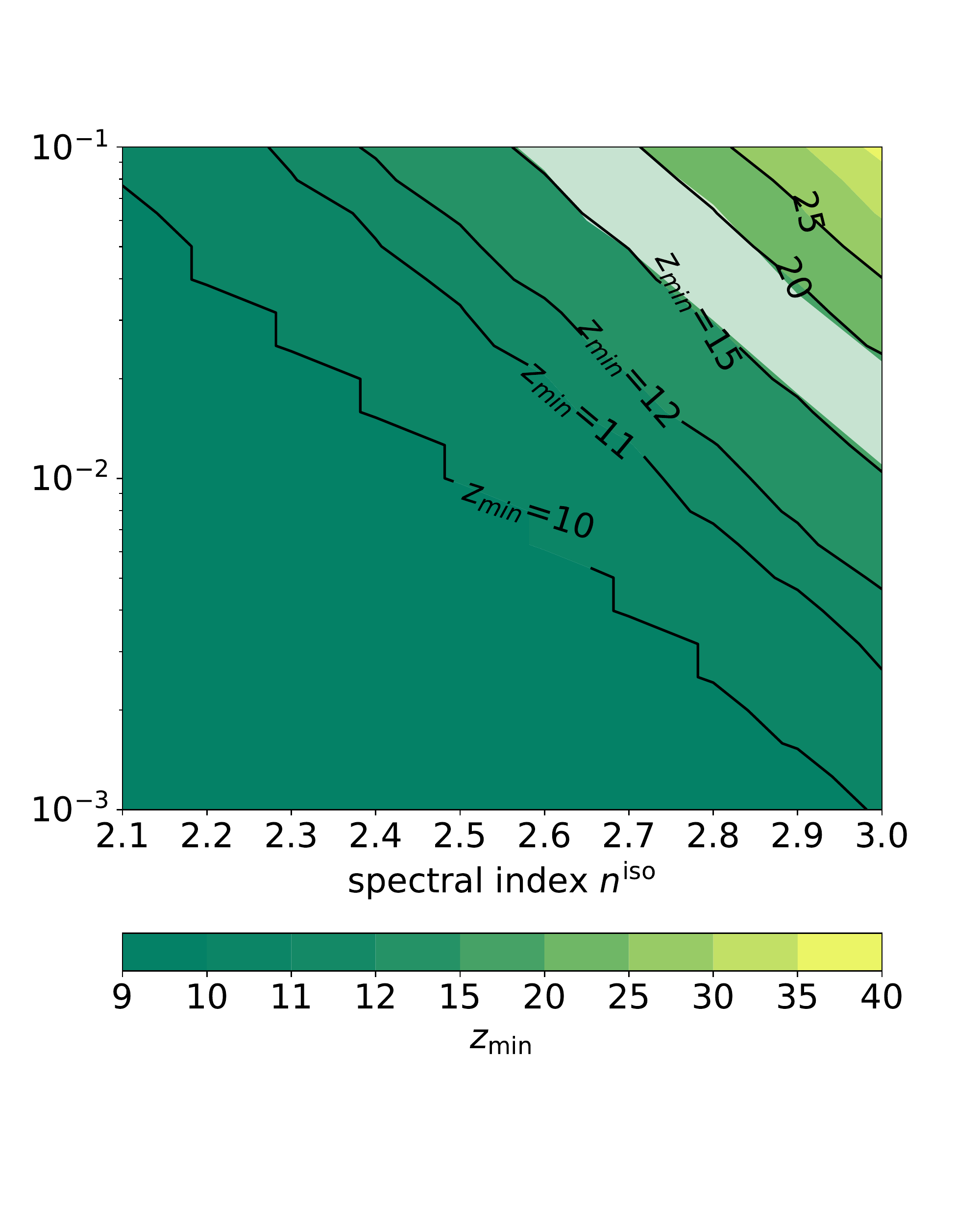}
  \end{minipage}
\caption{
Same as Figure \ref{fig:2dConstraint_m1}, but with the astrophysical model 2 (left), and model 3 (right).
}
\label{fig:2dConstraint_m23}
\end{figure*}

To clarify the uncertainties of astrophysical models in constraining isocurvature perturbations, we plot the absorption redshifts $z_\mathrm{min}$ calculated based on models 2 and 3 in Figure \ref{fig:2dConstraint_m23}. In astrophysical model 2, the position of the absorption trough is consistent with the EDGES observation of $15<z_\mathrm{min}<20$ without isocurvature fluctuations, and hence the constraints on the isocurvature perturbations are zero-consistent. On the other hand, the adiabatic case for model 3 produces the global signal with $z_\mathrm{min}=9.4$, as shown in Section \ref{sec:astromodels}. Therefore large isocurvature fractions are required when the absorption redshift is $15<z_\mathrm{min}<20$.

As shown in the left panel of Figure \ref{fig:2dConstraint_m23}, even if the isocurvature perturbations are less dominant (e.g. $n^\mathrm{iso}<2.5$ and $r_\mathrm{CDM}<0.01$), the isocurvature perturbations shift $z_\mathrm{min}$ to higher redshift while such weak isocurvature perturbations do not affect the signal much in the other two models. Thus the global signals with model 2 are more sensitive to the isocurvature parameters than other models. This is because  the isocurvature perturbations can enhance the number density of small halo and model 2 has a low $M_\mathrm{turn}$ and a low $\alpha_{*}$ which allows small halo to contribute to the {Lyman-$\alpha$} coupling at early epochs. This result infers a strong degeneracy between $M_\mathrm{turn}$ and the parameters of isocurvature perturbations. The degeneracy would be solved by observing the faint end of high-$z$ galaxy luminosity functions. 

\begin{figure}
\centering
\includegraphics[width=9cm]{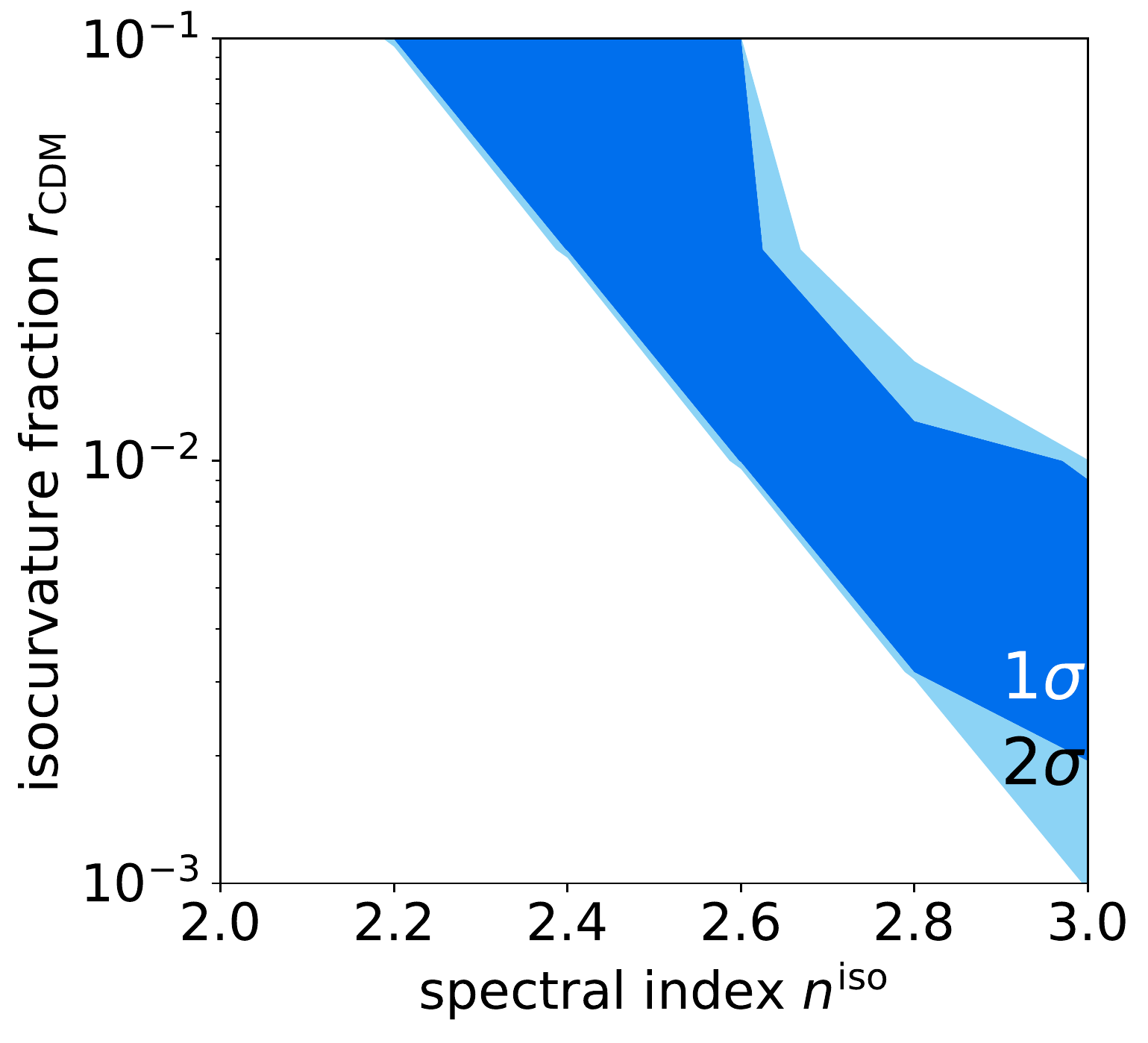}
\caption{
{1$\sigma$ (deep blue) and 2$\sigma$ (light blue) constraint in the $n^\mathrm{iso}-r_\mathrm{CDM}$ plane, which correspond to $\Delta \chi^2 \le 2.30$ and 6.18, respectively. Astrophysical parameters are marginalized to obtain the constraint.
}
}
\label{fig:chi2}
\end{figure}

{To end this section, we provide a further discussion on the degeneracy
among the astrophysica and isocurvature mode parameters.
In section~\ref{sec:astromodels}, we consider the impact of four astrophysical parameters $(\alpha_*,~ M_\mathrm{turn},~ t_*,$ $L_{X<2.0\mathrm{keV}}/\mathrm{SFR})$ on the 21-cm global signal.
Among these parameters, $M_\mathrm{turn}$ and $L_{X<2.0\mathrm{keV}}/\mathrm{SFR}$ are sensitive to the central position of the absorption profile $z_\mathrm{min}$.
Therefore, we perform the $\chi^2$ analysis with the isocurvature parameters $(r_\mathrm{CDM}, n^\mathrm{iso})$ and the astrophysical ones $(M_\mathrm{turn}, L_{X<2.0\mathrm{keV}}/\mathrm{SFR})$ as following:
\begin{enumerate}[(1)]
    \item For a given isocurvature parameter set $(r_\mathrm{CDM}, n^\mathrm{iso})$, we calculate the global signals with $5\times5$ astrophysical parameters, $M_\mathrm{turn}=$(1.58, 2.45, 3.80, 7.59, 14.8)$\times 10^8 M_\odot$ and $\log_{10} (L_{\mathrm{X<2.0keV}}/\mathrm{SFR}/[\mathrm{erg~s}^{-1} M_\odot^{-1}~\mathrm{yr}])=$(39.47, 40.05, 40.64, 41.08, 41.52), 
    which are varied within the 1$\sigma$ uncertainties reported by HERA \cite{2021arXiv210802263T}.
    \item We find the theoretical absorption redshift $z_\mathrm{min, th}$ for each parameter set, and estimate $\chi^2$ with the observational data as
    \begin{align}
    \chi^2 (\bm{p}) = \cfrac{(z_\mathrm{min,th}(\bm{p})-z_\mathrm{min,obs})^2}{{\Delta z_\mathrm{obs}}^2}~,
    \label{eq:chisqared}
    \end{align}
    where $\bm{p}\equiv(r_\mathrm{CDM}, n^\mathrm{iso}, M_\mathrm{turn}, L_{\mathrm{X<2.0keV}}/\mathrm{SFR})$ denotes the model parameters, and we assume $z_\mathrm{min,obs}=17.2$ and $\Delta z_\mathrm{obs} = 0.2$,
    which is translated from $\nu_\mathrm{center}=78\pm1~\mathrm{MHz}$ as EDGES reported absorption profile.
    \footnote{{The recent experiment by SARAS 3~\cite{2022NatAs.tmp...47S} excluded the absorption signal as EDGES obtained at almost 2 $\sigma$ C.L. However, the aim of this work is to discuss how severe constraint on the isocurvature parameters we can obtain from the current observational data.}}
    \item For each isocurvature parameter set, we choose the best astrophysical parameters so that the $\chi^2$ value in Eq.~\eqref{eq:chisqared} is minimized.
\end{enumerate}
Through these processes, we estimate 
$\Delta \chi^2 = \chi^2 - \chi_{\rm min}^2$ 
for different isocurvature parameters as shown in Figure \ref{fig:chi2}.
Here, the parameters giving the minimum $\chi^2$, denoted as $\chi^2_{\rm min}$, are $\log_{10} r_\mathrm{CDM}=-1.5$, $n^\mathrm{iso}=2.6$, $M_\mathrm{turn}=1.58\times 10^8 M_\odot$, $\log_{10} (L_{\mathrm{X<2.0keV}}/\mathrm{SFR}/[\mathrm{erg~s}^{-1} M_\odot^{-1}~\mathrm{yr}])=39.47$, and $\chi_{\rm min}^2=0.04$.
Figure \ref{fig:chi2} shows the difference from this best-fitted one for each parameter set.
The combined analysis of the EDGES global signal and 21-cm power spectrum with HERA give a tighter constraint on the blue-tilted isocurvature perturbations, roughly fitted by
\begin{align}
4.5 \le 2.5 n^\mathrm{iso}+\log_{10} r_\mathrm{CDM} \le 5.5
\hspace{2cm} (2 \sigma \text{ C.L.}).
\end{align}
The larger (or smaller) amplitude of isocurvature power spectra than the above range 
give the global signals with higher (lower) absorption redshifts.
}

\section{Conclusion}
In this work, we provide a new method to constrain the isocurvature perturbations with the 21-cm global signal, focusing on the blue-tilted CDM isocurvature fluctuations. We take two free parameters for the isocurvature perturbations, the isocurvature fraction $r_\mathrm{CDM}$ and the spectral index $n^\mathrm{iso}$. To discuss degeneracies between the astrophysical and isocurvature parameters, we use the galaxy-driven astrophysical model accommodated in the numerical calculation with \texttt{21cmFAST}. We vary several astrophysical parameters to control the star formation efficiency and X-ray heating under the constraint derived by the recent HERA observations and performed the numerical calculations.

We find that one can obtain the constraints on the isocurvature fluctuations from the observations of absorption redshifts of the 21-cm global signal. If lower absorption redshift $z_\mathrm{min}$ is observed, the constraint on the isocurvature gets tighter. Assuming that the absorption trough is observed with $z_\mathrm{min}<20.0$, the isocurvature fraction should be $r_\mathrm{CDM} \lesssim 0.001$ (model 2), 0.007 (model 1), $0.025$ (model 3) for $n^\mathrm{iso}=3.0$. By adopting the constraints on the astrophysical parameters obtained by HERA \cite{2021arXiv210807282T}, one can obtain a severe upper bound on $r_\mathrm{CDM}$.
{In addition, we performed the $\chi^2$ analysis to provide the constraint on the isocurvature perturbations by marginalizing the astrophysical uncertainties. When the absorption redshift and its error are given by $z_\mathrm{min}=17.2$ and $\Delta z=0.2$ as suggested by EDGES experiments, we obtain the constraint on the isocurvature parameters as $4.5 \le 2.5 n^\mathrm{iso}+\log_{10} r_\mathrm{CDM} \le 5.5 \ (2 \sigma ~\text{C.L.})$, under the astrophysical parameters constrained by HERA Phase-I.
}

From our results, large isocurvature fluctuations possibly produce a high redshift absorption signal as $z_\mathrm{min} \gtrsim 40$, even in the HERA-constrained astrophysical models. However, such early EoR scenarios should be excluded by other observational probes, such as the high-$z$ galaxy luminosity function and Thomson optical depth of the CMB photons. One can bring tighter constraints on the isocurvature fluctuations by combining those measurements and the 21-cm line observations. Such a combined analysis introduces another degeneracy between the isocurvature and astrophysical parameters, e.g., the escape fraction of the ionizing photons. This degeneracy would make the results more complicated, and it is beyond the scope of this paper. Therefore, we leave them as future work.

\section*{Acknowledgements}
T.M. is supported by JSPS Overseas Research Fellowships. S.Y. is supported by JSPS Research Fellowships for Young Scientists. The work was supported by JSPS KAKENHI Grant Number 17H01131 (T.T.),  19K03874 (T.T.), 21J00416 (S.Y.), and MEXT KAKENHI Grant Number 19H05110 (T.T.).
Numerical computations were carried out on Cray XC50 at the Center for Computational Astrophysics, National Astronomical Observatory of Japan.

\clearpage 
\bibliography{main}

\end{document}